\documentclass[twocolumn]{aa}  

\usepackage{graphicx}
\usepackage{txfonts}
\usepackage[colorlinks=true,urlcolor=blue,linkcolor=blue,citecolor=blue]{hyperref}
\usepackage{amsmath}	% Advanced maths commands
\usepackage{amsfonts}
\usepackage{mathrsfs}
\usepackage{subcaption}
\usepackage[retainorgcmds]{IEEEtrantools}
\usepackage{gensymb}
\usepackage{verbatim}
\usepackage{tikz}
\usepackage{tikz-3dplot}
\usepackage{siunitx}
\usepackage{color}
\usepackage{pdflscape}
\usepackage{commath}
\usepackage{xcolor}
\usepackage{caption}
\usepackage{ulem}

\begin{document} 
\title{Accretion discs onto supermassive compact objects: a portal to dark matter physics in active galaxies}
% Accretion discs in alternative theories of supermassive black holes: a portal to dark matter physics
% Accretion discs on to supermassive black hole alternatives: dark matter fermion cores in active galaxies
% Accretion discs on to horizonless compact objects: the case of dark matter fermion cores in active galaxies
%   \subtitle{}
\titlerunning{Accretion discs onto supermassive compact objects}
   \author{
    C. Millauro\inst{1}
    \and
    C. R. Arg\"uelles\thanks{E-mail: carguelles@fcaglp.unlp.edu.ar} \inst{2,3}
    \and
    F. L. Vieyro \inst{4,5}
    \and
    V. Crespi \inst{5}
    \and
    M. F. Mestre \inst{3,5}
    }
% List of institutions
    \institute{Departamento de Física, Facultad de Ciencias Exactas y Naturales, Universidad de Buenos Aires, Pabellón I, Ciudad Universitaria, 1428 Buenos Aires, Argentina \\
    \and
    ICRANet, Piazza della Repubblica 10, 65122 Pescara, Italy\\
    \and
    Instituto de Astrofísica de La Plata, UNLP $\&$ CONICET, Paseo del Bosque, B1900FWA La Plata, Argentina\\
    \and
    Instituto Argentino de Radioastronomía (IAR, CONICET/CIC/UNLP), C.C.5, (1894) Villa Elisa, Buenos Aires, Argentina \\
    \and
    Fac. de Ciencias Astron. y Geof\'isicas, Universidad Nacional de La Plata, Paseo del Bosque, B1900FWA La Plata, Argentina
    }

   \date{Received XX, XX; accepted XX, XX}

% \abstract{}{}{}{}{} 
% 5 {} token are mandatory
 
  \abstract
{The study of the physics of accretion discs developed around the supermassive black hole (BH) candidates are essential theoretical tools to test their nature.}
{Here, we study the accretion flow and associated emission using generalised $\alpha$-discs on to horizonless dark compact objects, in order to compare with the traditional BH scenario. The BH alternative here proposed consists in a dense and highly degenerate core made of fermionic dark matter (DM) which is surrounded by a more diluted DM halo. Such a \textit{dense core}--\textit{diluted halo} DM configuration is a solution of the Einstein equations of General Relativity (GR) in spherical symmetry, which naturally arises once the quantum nature of the DM fermions is dully accounted for.}
{The methodology followed in this work consist in first generalising the theory of $\alpha$-discs to work in the presence of regular and horizonless compact objects, and second, to apply it to the case of \textit{core}-\textit{halo} DM profiles typical of active-like galaxies.}
 {The fact that the compactness of the dense and transparent DM core scales with the particle mass, allows for the following key findings of this work: (i) it always exist a given core compacity -i.e. corresponding particle mass- which produces a luminosity spectrum which is basically indistinguishable from that of a Schwarzschild BH of the same mass as the DM core; (ii) the disc can enter deep inside the non-rotating DM core, allowing for accretion powered efficiencies as high as $28\%$, thus comparable to that of a highly rotating Kerr BH. }
{These results, together with the existence of a critical DM core mass of collapse into a supermassive BH, open new avenues of research for two seemingly unrelated topics such as AGN phenomenology and dark matter physics.}


   \maketitle
%
%-------------------------------------------------------------------

\section{Introduction}
%las observaciones cosmológicas obtenidas en las últimas tres décadas han favorecido la adopción del paradigma $\Lambda$CDM \citep{bahcall1999,ivanov2020}
One of the central results of the standard cosmological model {\sl Lambda cold dark matter} ($\Lambda$CDM), is the necessity of invoking a dark matter component as an essential part in the composition of the Universe \citep{bahcall1999}. However,  how such a dark component of matter is distributed on inner galactic scales, and what is the nature and mass of the dark matter particles, still remain unclear \citep{2017ARA&A..55..343B}. A main available tool to tackle these questions is based on cosmological N-body (classical) simulations with adequate initial conditions, as for example the ones provided by the $\Lambda$CDM paradigm (see e.g. \citealt{ivanov2020} and references therein). Although this paradigm manages to explain in a good way the distribution of dark matter on large scales ($>$ Mpc), it faces various challenges on short galactic scales \citealt{2005MNRAS.364..665D,2008ApJ...681L..13B,2009ApJ...692L...1J,2017ARA&A..55..343B}). 

Within the framework of cosmological simulations, different state-of-the-art alternatives are being provided to try to solve these problems, including the possibility that cold DM is self-interacting \citep{2020JCAP...06..027K}, considering warm instead of cold DM \citep{2019MNRAS.483.4086B}, or even abandoning the  hypothesis of classical particles by incorporating quantum effects into the simulations \citep{2014NatPh..10..496S}. Along the line of including the quantum nature of the DM particles explicitly in the physics of the DM halos, it has been recently proposed an alternative (semi-analytical) approach for fermionic DM in a cosmological framework \citep{2021MNRAS.502.4227A}. It covers the problems of DM halo formation, overall morphology and stability, from first principle physics. In particular, it includes for (quantum) statistical mechanics and thermodynamics in the presence of self-gravity, offering solutions to some of the problems that the $\Lambda$CDM paradigm possesses at short scales (see e.g., \citealt {2023ApJ...945....1K,2023Univ....9..197A} for a recent work and a review, respectively).

Fermionic mass distributions of this sort are obtained by solving the equations of a self-gravitating system of neutral fermionic (spin $1/2$) particles in hydrostatic and thermodynamic equilibrium in GR. Some generic solutions to this model were obtained in the past in different works aimed to the problem of DM halos \citep{1984ApJ...281..560C,1988NCimB.101..369I,1990A&A...235....1G,1998MNRAS.296..569C,2002PrPNP..48..291B,2006IJMPB..20.3113C,2013NewA...22...39D,2014IJMPD..2342020A,2015MNRAS.451..622R,2015PhRvD..92l3527C}, though only recently a more realistic version of this theory including for particle evaporation and central (fermion) degeneracy, was developed in GR in \citet{arguelles2018,arguelles2019,2021MNRAS.502.4227A} and is referred to as the (extended) Ruffini-Arg\"uelles-Rueda (RAR) model\footnote{The term `relativistic fermionic King model' is also used in the literature \citep{2022PhRvD.106d3538C}.}. The model implies novel DM density profiles which self-consistently accounts for the Pauli exclusion principle, thus yielding a source of quantum pressure towards the centre of the configurations with key implications for galactic nuclei. The more general DM profiles develop a \textit{dense core} - \textit{diluted halo} morphology which, unlike other phenomenological profiles in the literature, depends on the mass of the particle. Remarkably, these fermionic DM halos can explain the galaxy rotation curves in different galaxy types \citep{arguelles2018,arguelles2019,2023ApJ...945....1K}, while the degenerate fermion core can mimic their central BHs \citep{arguelles2018,arguelles2019,bec2020,2021MNRAS.502.4227A,bec2021,2022MNRAS.511L..35A,2022IJMPD..3130002A}. Moreover, as demonstrated in \cite{2021MNRAS.502.4227A,2023MNRAS.523.2209A,arguelles2023c} from dynamical and thermodynamical stability criteria in GR, the central DM core can reach a critical mass for collapse and thus providing a novel channel for supermassive BH formation in the early Universe (see also \citealt{chavanis2019,chavanis2020} for a first dynamical and thermodynamical instability study in GR of the self-gravitating Fermi gas at finite temperature leading to a BH formation).%\citep{2023MNRAS.523.2209A}.

In this work we will centre our attention on fermionic \textit{core}-\textit{halo} profiles applied to typical active-like galaxies, together with their central accretion processes. The study of the free parameters of the theory (including particle mass) will be focused in solutions whose central core has not yet reached the critical mass for collapse, and thus it will represent an alternative to the traditional BH scenario. This choice is being motivated by the ambitious endeavour of trying to understand the very nature of the massive compact objects at galaxy centres, their formation channel, surrounding emission, and finally their relation with the host galaxy and AGN phenomenology. %As it will be clear from this work, the model of fermionic DM halos together with the disc accretion physics here applied can shed light to all of the above questions.  

Motivated by distinct branches of theoretical physics and astrophysics, different alternative models to that of the classical BHs have been proposed to date (see e.g. \citealt{2019LRR....22....4C} for a review). When dealing with galaxy centres, a typical example studied in the recent past are boson stars \citep{1999GReGr..31..787S,2000PhRvD..62j4012T,2006PhRvD..73b1501G,vincent2016,olivares2020}, that is, horizonless and massive compact objects made of self-gravitating scalar fields. In particular, different observational signatures of boson stars were studied, such as the luminosity spectra of $\alpha$-discs \citep{2006PhRvD..73b1501G};  strong-field images and luminosity patterns in boson stars surrounded by a disc torus \citep{vincent2016}; and a detailed study of the accretion flow via general relativistic magnetohydrodynamic simulations in the space-time of a boson star \citep{olivares2020}.  

In analogy to the above cases of study, it is our objective to start a research program for AGN phenomenology dedicated to the RAR model for a self-gravitating system of fermions as DM in galaxies. Thus, in order to cover the main observational signatures associated to the emission of galaxy centres, we start in this work by studying the accretion and corresponding luminosity of barionic matter onto supermassive compact cores made of fermionic DM. This will be done by first extending the standard disc model of Shakura \& Sunyaev \citep{shakura1973} in presence of a fermionic DM distribution, using a Keplerian disc and a classical treatment. In the case of fermionic particles, the main motivation comes via the numerous efforts made in the last decade to shed light on the nature of such supermassive dark compact objects and the surrounding DM halo in a unified description \citep{2023Univ....9..197A}. 

With the data coming from the observational campaigns dedicated to the stellar motions around Sgr A* \citep{ghez2005,ghez2008,genzel2010,grav2018} -confirming the presence of a supermassive compact object- it has been shown that the core-halo RAR solutions accurately reproduce the orbital motion of the S stars including its relativistic effects \citep{bec2020,bec2021,arguelles2022sga}. Additionally, the observations of the relativistic images using Very Long Base Interferometry (VLBI) in both M87 and the Milky Way \citep{collaboration2019first,akiyama2022first}, motivate to further test the RAR solution for which it is essential to study the accretion physics in this new paradigm.
%(fulfilling with Pauli principle and particle evaporation)

The article is organised as follows: we briefly describe the extended RAR model in Sec. \ref{rarmodel}. In Sec. \ref{accretion}, we study the efficiency, spectra as well as solutions of steady-state thin discs which accrete in the background metric of the RAR model, and in Sec. \ref{conclusions} we present the conclusions of the work. 

% As a first step of the above research program, we start by studying the accretion of barionic matter onto suppermasive compact cores made of fermionic DM, the later according to the extended RAR model.
%From a physical point of view, accretion discs are classified into different regimes according to the accretion rate of matter on the compact object [7]. For high accretion rates, the flow can be modeled with the Shakura-Sunyaev thin and optically thick disc standard solution (SSD, [8.9]). These discs are characterized by extending to the last stable orbit when they form around black holes. In the case of the RAR model, due to the regularity of the central object, there is no innermost stable circular orbit. This will be treated in section  \ref{rarmodel}. 

%--------------------------------------------------------------------
\section{Extended RAR model}\label{rarmodel}

\subsection{The model}

The RAR model consists on a spherical system of self-gravitating tempered fermions which are distributed in phase-space according to the following Fermi–Dirac-like distribution function:
\begin{equation}\label{Eq:1}
    f_c(\epsilon \leq \epsilon_c) = \frac{2}{h^3}\frac{1-e^{(\epsilon-\epsilon_c)/kT}}{e^{(\epsilon-\mu)/kT}+1}, ~~~~~~~~ f_c(\epsilon > \epsilon_c) = 0.
\end{equation}

\noindent With $\epsilon = \sqrt{c^2p^2+m^2c^4}-mc^2$ the particle kinetic energy, $\mu$ is the chemical potential with the particle rest-energy subtracted off, $T$ is the effective temperature, $k$ is the Boltzmann constant, $h$ is the Planck constant, $c$ is the speed of light, and $m$ is the fermion mass. Anti-fermions are not included as temperatures $T<<mc^2/k$ are considered. The full set of (functional) dimensionless-parameters of the model is defined by the temperature, degeneracy and cutoff parameters, $\beta = kT/(mc^2)$, $\theta = \mu/kT$, $W=\epsilon_c/(kT)$ respectively.

Interestingly, a coarse-grained phase-space distribution of this kind can be linked with halo formation processes, since it can be obtained from a generalised kinetic theory in presence of gravity as demonstrated in \cite{2004PhyA..332...89C}. Indeed, it was shown there that Eq. \ref{Eq:1} is a (quasi-) stationary solution of a generalised Fokker-Planck equation for fermions \citep{2004PhyA..332...89C,2006IJMPB..20.3113C}. Such a kinetic theory includes the physics of violent relaxation appropriate for nonlinear structure formation -as originally presented in \cite{1967MNRAS.136..101L} for classical particles- though further extended including for particle evaporation and applied to realistic DM halos \citep{2015PhRvD..92l3527C,2021MNRAS.502.4227A}. This kind of phase-space distributions have been shown to fulfil a maximisation entropy principle during the collisionless relaxation process, until the halo reaches the steady state which is currently observed. More recently, this formation mechanism of fermionic halos was applied in \cite{2023ApJ...945....1K} for a sample of $120$ galaxies, and compared with phenomenological profiles as obtained from cosmological N-body simulations. 

The corresponding 4-parametric fermionic equations of state, at given radius $r$, $\rho(\beta, \theta , W, m)$, $P(\beta, \theta , W, m)$, are directly obtained as the corresponding integrals (bounded from above by $\epsilon \leq \epsilon_c$ ) of $f_c(p)$. These components are the diagonal part of the stress–energy tensor in the Einstein equations, which are solved under the perfect fluid approximation within a background metric with spherical symmetry, which reads
\begin{equation}
    ds^2 = e^\nu c^2dt^2 - e^\lambda dr^2-r^2d\Theta^2-r^2sin^2\Theta d\phi^2,
\end{equation}
with $(r, \Theta, \phi)$ the spherical coordinates, and $\nu$ and $\lambda $ only depending on the radial coordinate $r$. The system of Einstein equations (that is mass and TOV equations below) is solved together with the Tolman and Klein thermodynamic equilibrium conditions (involved in Eqs. \ref{eq:RAR3}-\ref{eq:RAR4} below), and (particle) energy conservation along a geodesic (Eq. \ref{eq:RAR5} below). The dimensionless system of highly-nonlinear ordinary differential equations reads:
\begin{equation}\label{eq:RAR1}
    \frac{d\hat{M}_{DM}}{d\hat{r}} = 4\pi\hat{r}^2\hat{\rho},
\end{equation}
\begin{equation}\label{eq:RAR2}
    \frac{d\nu}{d\hat{r}} = \frac{2(\hat{M}_{DM}+4\pi\hat{P}\hat{r}^3)}{\hat{r}^2(1-2\hat{M}_{DM}/\hat{r})},
\end{equation}
\begin{equation}\label{eq:RAR3}
    \frac{d\theta}{d\hat{r}} = -\frac{1-\beta_0(\theta-\theta_0)}{\beta_0}\frac{\hat{M}_{DM}+4\pi\hat{P}\hat{r}^3}{\hat{r}^2(1-2\hat{M}_{DM}/\hat{r})},
\end{equation}
\begin{equation}\label{eq:RAR4}
    \beta(\hat{r}) = \beta_0e^{\frac{\nu_0-\nu(\hat{r})}{2}},
\end{equation}
\begin{equation}\label{eq:RAR5}
    W(\hat{r}) = W_0 + \theta(\hat{r}) - \theta_0,
\end{equation}
where the dimensionless quantities are: $\hat{r} = r/\chi$, $\hat{M}_{DM} = GM_{DM}/(c^2\chi)$, $\hat{\rho}=G\chi^2\rho/c^2$, $\hat{P} =G\chi^2P/c^4$, with $\chi = 2\pi^{3/2}(\hbar/(mc))(m_p/m)$ and $m_p = \sqrt{\hbar c/G}$ the Planck mass. The system of Eqs. \ref{eq:RAR1}-\ref{eq:RAR5} constitute a boundary condition problem which, for fixed DM particle mass $m$, has to be solved for given set of free parameters ($\beta_0$,$\theta_0$,$W_0$) defined at the centre of the configuration.     

The most general solution results in a degenerate compact core (governed by Pauli degeneracy pressure), surrounded by an extended and more diluted halo (governed by thermal pressure) as detailed in section below and in \cite{arguelles2018}. The core mass $M_c = M_{DM}(r_c)$ is given at the core radius, defined as the first maximum of the rotation curve. It  corresponds with the radius where the central density has decreased about one tenth of the central value, i.e. where fermion degeneracy starts to vanish. In the following section we show different solutions for well motivated values of $m$, having DM core masses and total halo masses typical of active-like galaxies following \cite{arguelles2019}. 

\subsection{Application to active-like galaxies}

\begin{figure}
    \centering
    \includegraphics[width=\columnwidth]{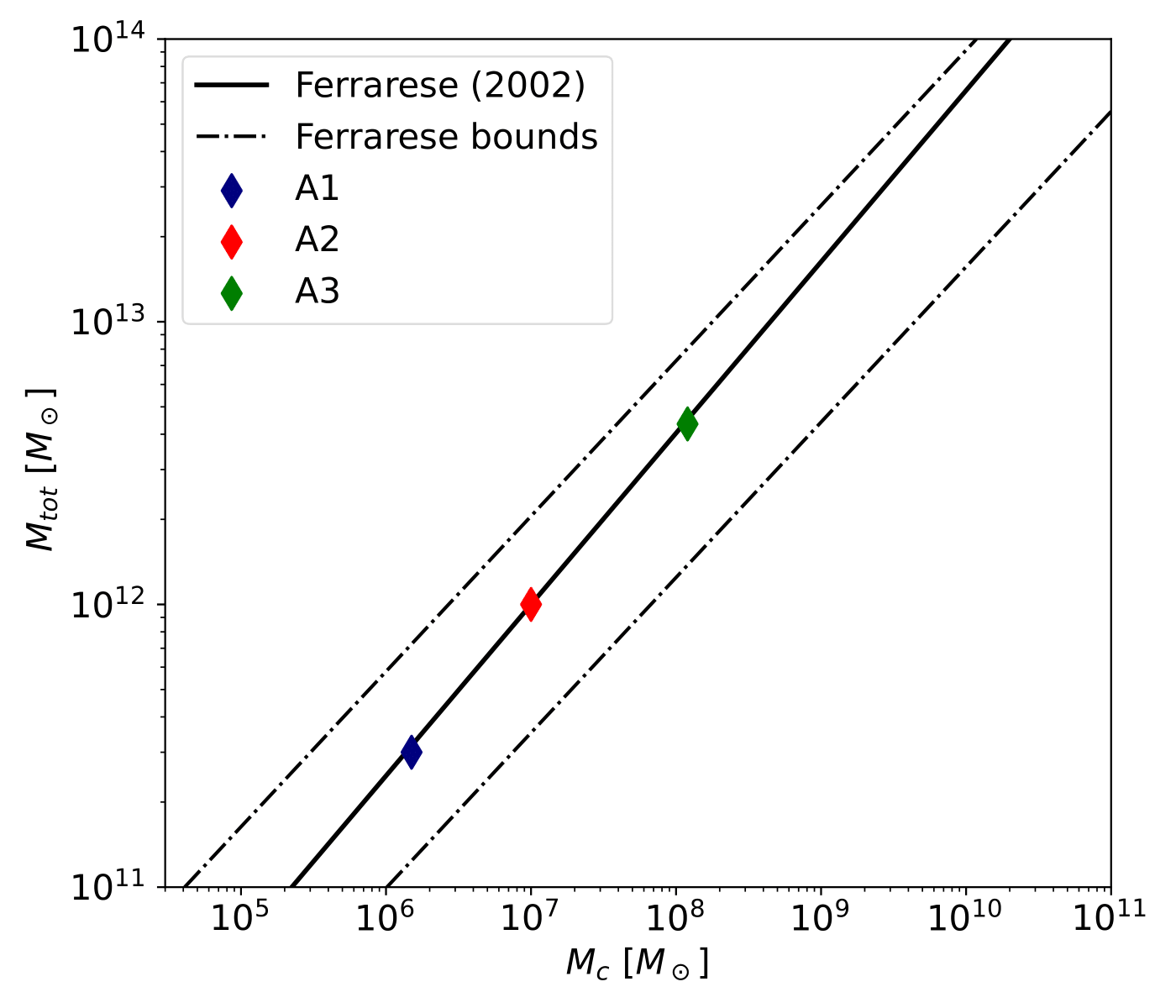}
    \caption{\textit{core}-\textit{halo} RAR solutions for DM particle mass of $mc^2=48$ keV, in agreement with the Ferrarese relation connecting the halo mass with the supermassive central object mass \citep{ferrarese2002}.}
    \label{fig:Ferrarese}
\end{figure}

The application of the extended RAR model to large galaxies with dark and regular compact 
cores reaching mass values of $\sim 10^7$ - $10^8 M_\odot$ typical of AGN, was first shown in \cite{arguelles2019}. 
This was done for a DM particle mass $m \approx 50$ keV motivated by the excellent results obtained for the Milky Way \citep{arguelles2018,bec2020,bec2021,arguelles2022sga}, where the corresponding DM core explains the motion of the S-cluster stars around Sgr A*. 
For such value of the fermion mass, the highly degenerate core reaches the critical mass of collapse $M_c^{cr}$ into a SMBH of $\approx 2\times 10^8 M_\odot$ \citep{2021MNRAS.502.4227A}. 
%\sout{Larger BH masses can be further obtained by disc accretion within standard Eddington-like rates }\citep{2023MNRAS.523.2209A}, \sout{with key implications to SMBH formation in the high $z$ Universe}. 
Particle masses of the order of $\sim 100-350$ keV have also been analysed within this DM model with excellent results (see \citealt{arguelles2018,2023MNRAS.523.2209A}). 
As demonstrated in \cite{2014IJMPD..2342020A,2021MNRAS.502.4227A} the larger the $m$ the  lower the critical core masses $M_c^{cr} \approx M_{\rm OV}\propto 1/m^2$ (roughly following the Oppenheimer-Volkoff limit \citealt{1939PhRv...55..374O}), with $m\approx 350$ keV leading to a critical mass of a DM core collapsing to a SMBH of $\approx 4\times 10^6 M_\odot$ as for SgrA* \citep{arguelles2018,bec2020}.

An important prediction of the \textit{core}-\textit{halo} family of RAR DM profiles for given $m$, is that it can answer for different universal scaling relations such as: the Ferrarese relation \citep{ferrarese2002, 2015ApJ...800..124B} connecting the halo and its supermassive central object masses; the DM surface density relation \citep{2009MNRAS.397.1169D}, and the radial acceleration relation \citep{2016PhRvL.117t1101M}; as shown in \citep{arguelles2019,2023ApJ...945....1K}. Indeed, in Fig. \ref{fig:Ferrarese} we show an example of RAR profiles with  $m=48$ keV (solutions $A1-A3$), corresponding to a halo mass window $M_{\rm tot}\sim 10^{11}-10^{12} M_\odot$ and supermassive DM compact objects of mass $M_c\sim 10^6 - 10^8 M_\odot$ in excellent agreement with the Ferrarese relation (RAR models with $m=200$ keV are also in agreement with the correlation though reaching up to $M_c\sim 10^7 M_\odot$).
 
%[Carlos: Here is must be clear and easy to understand for the general reader, why the RAR model is said to be applied to active galaxies. First it has to be clearly detailed the main results of PDU II regarding that, for fixed $mc^2\approx 48$ keV, it exist a family of RAR profiles able to explain the DM halos from dwarf to big elliptical and galaxy clusters, while predicting the existence of dense DM cores at their centers, which, in the case of large enough galaxies they follow the Ferrarese relation. At this point I believe it should be important to add a new Figure (or at least put in the text the Mc-Mtot relation equation!, saying that the two typical solutions here considered (i.e. for the two particle masses) lie along this relation!). No much emphases to the MW case must be given, but just on how we motivate the choice of 48 keV mass.]

\begin{figure}
    \centering
    \includegraphics[width=\columnwidth]{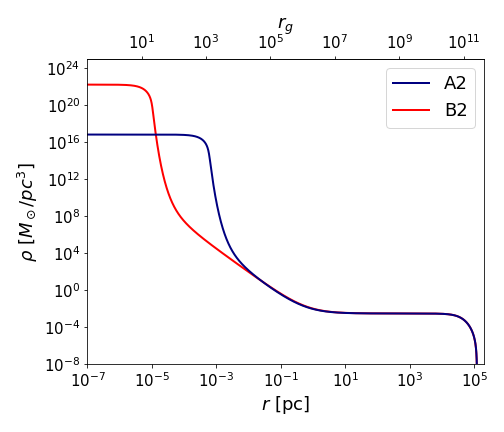}
\caption{Density profiles corresponding to core mass of $M_{\rm c} = 1.0\times 10^7 M_{\odot}$, and particle masses of $mc^2=48$ keV (A2, blue) and $mc^2=200$ keV (B2, red). These DM halos correspond to typical Elliptical galaxies.}
\label{fig:m-rho}
\end{figure}

\begin{figure}
    \centering
    \includegraphics[width=\columnwidth]{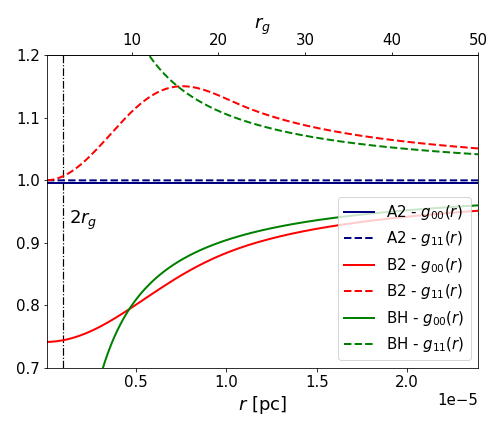}
\caption{Metric components corresponding to the same RAR solutions as in Fig. \ref{fig:m-rho} with core mass of $M_{\rm c} = 10^7 M_{\odot}$, and particle masses of $mc^2=48$ keV (A2, blue) and $mc^2=200$ keV (B2, red). A comparison with the metric of a Schwarzschild BH of mass $M_{\rm BH} = 10^7 M_\odot$ is also shown (BH, green).}
\label{fig:metrics}
\end{figure}

One of the central objectives of this work is to analyse the accretion power efficiencies, and consequent luminosity spectrum, caused by supermassive compact objects alternative to BHs. %\sout{Different DM core-compacities for given core mass will imply appreciable differences on the above magnitudes with respect to the traditional BH scenario, which are worth of a detailed study. Indeed,}
Due to the fact that larger fermion masses imply more compact and denser DM cores at given core mass (see e.g. \cite{arguelles2018}), we will then study here two different families of profiles: the ones with $mc^2= 48$ keV (which we call RAR model A), and the ones with $mc^2 = 200$ keV (referring to the RAR model B). In Table \ref{table_1} we show the main parameters of both models: for a typical DM compact-core mass of $10^7 M_\odot$, the model B2 gives a more compact core with respect to the A2 solution (see Fig. \ref{fig:m-rho} for a comparison between both profiles). The DM core mass in solution B2 is close to the critical mass of gravitational collapse to a Schwarzschild BH of that mass and thus implying very similar metric functions (see Fig. \ref{fig:metrics}).

%the later having a DM core compacity closer to that of a Schwarzschild BH (see Fig.\ref{fig:metrics}). The values of the main parameters for the different models are listed on Table \ref{table_1}. The values of the compact core and the halos were fitted in order to satisfy the correlations given by \citet{ferrarese2002}.
% which is the less compact solution that reproduces the (circular averaged) orbital motion of the best resolved S-cluster stars \citep{arguelles2018}
%and , which provides the maximum (critical) core mass of $M_c^{cr} = 1.9 \times 10^8 M_{\odot}$. Hence,  all compact core studied in this work have masses below this critical value. The  other three free parameters of the model are adopted as (REF):
%\begin{enumerate}
%    \item $\theta_0 = 3.96\times 10^{1}$ 
%    \item $W_0 = 6.44\times 10^{1}$
%    \item $\beta_0 = 6.7\times 10^{-3}$
%\end{enumerate}

\begin{table*}%[ht]
    \caption[]{Main parameters of the different RAR models.}
   	\label{table_1}
   	\centering
\begin{tabular}{lccccccc}
\hline\hline %
Model & Particle mass  & $M_{\rm c}$ & $M_{\rm tot}$ & $r_{\rm c}$ &$\theta_0$ & $W_0$ & $\beta_0$ \\ [0.01cm]
 & [keV]  & [$M_{\odot}$] & [$10^{11}M_{\odot}$]  & [$r_{\rm g}$]& &&\\ [0.01cm]
\hline
A1 & 48     &$1.5 \times 10^6$ & 3 & $1.5\times 10^4$& 39.5& 69.6& $2.5\times 10^{-6}$\\ [0.01cm]
A2 & 48     &$1.0 \times 10^7$ & 10 & $1.0\times 10^3$& 37.9& 66.6&$6.8\times 10^{-5}$\\ [0.01cm]
A3 & 48     &$1.2 \times 10^8$ & 45 & $3.4\times 10^1$& 38.8& 67.2& $1.1\times 10^{-3}$\\ [0.01cm]

B1 & 200    &$3.5 \times 10^6$ & 3 & $8.1 \times 10^1$&49.4& 77.9& $5.1\times 10^{-3}$\\ [0.01cm]
B2 & 200    &$1.0 \times 10^7$ & 10 &  $1.5 \times 10^1$& 44.3& 75.4& $2.1\times 10^{-3}$\\ [0.01cm]

\hline\hline %
%\hline  \\[0.005cm]
\label{tab: 1}
\end{tabular}
\end{table*}

In Fig. \ref{fig:m-rho} we show the density profiles for models A2 and B2. They clearly show the existence of compact and massive cores corresponding with the highest density trends, and the diluted halo at larger radii. This is because the system goes from being governed by fermionic degeneracy pressure at small radii, to be governed by thermal pressure at larger radii. Thus, the density profile transitions from a degenerate core of almost constant density to a Boltzmann-like regime in the outer halo, where the density falls off as a power-law followed by an exponential decrease determining the galaxy border.
%Then it decays to zero due to the presence of the cutoff in the energy distribution. %In the case of the mass profile, it can be seen that from the radius of the nucleus the mass is constant for several orders of magnitude, and then it grows again in the middle of the region where the density is constant. 

%\subsection{Particle motion in the RAR-solution}

\section{Accretion discs}\label{accretion}
\subsection{Efficiency}\label{3.1}

There is an essential difference between the Schwarzschild BH solution and the RAR DM model regarding the motion of massive particles in their surroundings. Because the former is a singular solution of the Einstein equations and the latter is not,  the existence of the \textit{Innermost Stable Circular Orbit} (ISCO) becomes manifested. Due to the regularity of the central object in the RAR case, there is no critical angular momentum of the particle for which a potential barrier is no longer present, and therefore no ISCO can be reached \citep{crespi2022estudio}. This behaviour directly implies that there is no clear inner boundary as to where the disc might extend\footnote{General Relativistic simulations of fluid dynamics should be performed to tackle this problem in analogy to boson stars either for unmagnetised \citep{2016CQGra..33o5010M} or magnetised scenarios \citep{olivares2020}, though both are out of the scope of this paper.}. To set an inner constraint in this regard, we study the binding energy of a test particle in the disc, and analyse when it saturates towards a maximum.

\begin{figure}%[H]
    \centering
    \includegraphics[width=\columnwidth]{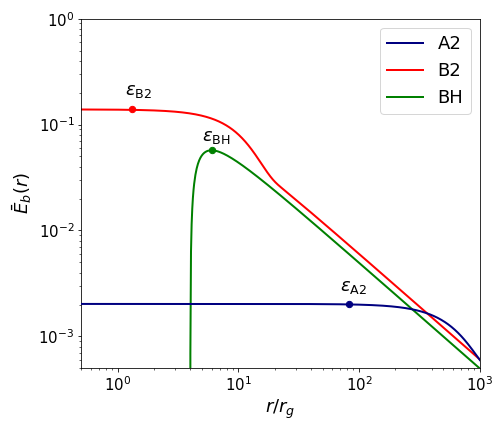}
    \caption{Accretion efficiencies (labelled with coloured dots) for two different DM core compacities (i.e. different $m$) having the same core mass of $M_{\rm c}=  10^7 M_{\odot}$ (RAR solutions $A_2$ and $B_2$). For comparison, the case of a Schwarzschild BH with a mass equal to $M_c$ is also shown. Interestingly, solution $B_2$ reaches an accretion efficiency of $14\%$, considerably larger than that of the Schwarzschild BH.}%\textcolor{blue}{(un grafico de este estilo pero con las masas de 48, 200 y 345; poner r en unidades de M o rg/rsch (idem todas las figuras))}}
    \label{fig:binding}
\end{figure}

In the case of BHs, the efficiency of the accretion process is related to the binding energy per unit rest mass at the innermost stable circular orbit (ISCO), that is:

\begin{equation}
    \varepsilon = \frac{mc^2 -E_{\rm c}}{mc^2}%=1-\sqrt{A(r)\left(1+\frac{r_cA'(r_c)}{2A(r_c)-r_cA'(r_c)}\right)}.
\end{equation}

\noindent where $E_{\rm c}$ corresponds to the particle's energy at the last stable circular orbit.
This accretion efficiency represents the maximum fraction of the rest mass energy of the accreted particle that can be converted into radiated energy. In standard astrophysical scenarios such as a neutron star of radius $\sim 10$ km, the efficiency is of order $10\%$, while in the case of BH accretion the efficiency varies between $5.7\%$ and $42\%$ for Schwarzschild and Kerr (maximal rotation with prograde disc) solutions respectively \footnote{When considering the effects of the radiation of the disc into the rotation of the black hole, the maximum spin reduces, and the efficiency results in $32.4\%$ \citep{thorne1974,laor1989}.} \citep{novikov1973}. In analogy to the BH case, and in order to define the corresponding accretion efficiency in the RAR model, we seek for a maximum in the radial behaviour of the (normalised) binding energy of a test particle. Thus, in the RAR-solution, the binding energy normalised with the rest mass reads:

\begin{equation}\label{eq:binding-energy}
    \bar{E}_{\rm b}(r) = 1-\sqrt{g_{00}(r)\left(1+\frac{rg_{00}'(r)}{2g_{00}(r)-rg_{00}'(r)}\right)}.
\end{equation}

\noindent where $g_{00}(r) = e^{\nu(r)}$ is the temporal component of the metric, and the prime symbol indicates derivation with respect to $r$. We show this binding energy behaviour in Fig. \ref{fig:binding} for the case of a Schwarzschild BH -obtained by replacing $g_{00}(r)$ in Eq. \ref{eq:binding-energy} by $1-2GM/r$-, and for two different RAR solutions with the same DM core mass of $M_{\rm c}= M_{\rm BH} = 10^7 M_{\odot}$ but different core-compacities. For the RAR solutions the binding energy asymptotically reaches the maximum as $r \rightarrow 0$, and saturates (see definition below) at a given radius $r_{\rm in}$. The definition for $r_{\rm in}$ implies that it is always smaller than the core radius, with the binding energy remaining approximately constant as the particle's orbit gets smaller and smaller (i.e. until $r_{\rm min} \sim 10^{-7}r_{\rm {g}} \approx 10^{-13}$ pc, the smallest representable radius admitted by computer precision for such \textit{core}--\textit{halo} solutions).
We adopt as the inner radius of the disc, $r_{\rm in}$, the value at which the relative error for the change in the maximum efficiency of the binding energy is equal or lower than $1\%$, that is:
\begin{equation}
    \frac{|\bar{E}_{\rm b}(r_{\rm in})-\bar{E}_{\rm b}(r_{\rm min})|}{\bar{E}_{\rm b}(r_{\rm min})} \leq 0.01
\end{equation}
Adopting this definition, the inner radius corresponds typically to a tenth of the core radius $r_{\rm in} \sim 0.1 r_{\rm c}$ (for $1\%$ of relative error). We compare the efficiencies of the RAR models $A_2$ and $B_2$ (labelled with $\varepsilon_{A2}$ and $\varepsilon_{B2}$ in Fig. \ref{fig:binding}), with the Schwarzschild one labelled with $\varepsilon_{\rm BH}$. Remarkably, since the disc can enter deep inside the DM core (that is the Keplerian orbits in the disc can have $r<r_c$), the binding energy of RAR solutions with compact enough cores can surpass the maximum Schwarzschild value. In Fig. \ref{fig:binding} we show an example for a below-critical DM core (solution $B_2$) with an efficiency of $14\%$, to be compared with the $5.7\%$ of the BH case. Moreover, when the degenerate core achieves its critical mass of collapse the accretion efficiency can be as large as $\approx 28\%$, similar to highly rotating Kerr BH. This interesting result is not new, and analogous conclusions were already obtained in the past for relativistic clusters with constant density and different compacities by \citep{1995NCimB.110...95C}.

These novel conclusions, exemplified in Fig. \ref{fig:binding}, have potential implications in the astrophysics of AGNs.
%, and its relevance deserves further explanation. 
In the context of the Soltan argument \citep{1982MNRAS.200..115S}, and based on observational results about the mean accretion efficiency of SMBHs (both from the local and the AGN-relic Universe, \citealt{2002MNRAS.335..965Y,2003ApJ...598..886U,2004MNRAS.351..169M}), it arises a mean value of $\epsilon\approx 0.1$ falling between the one predicted by the Schwarzschild and Kerr BH cases. Additionally, data analysis of different populations of AGNs shows that required efficiencies are $\epsilon >0.1$ \citep{2009MNRAS.396.1217R}, with the bulk of the cases having $\epsilon \sim 0.2$-$0.3$ and a minority of them reaching even down to $0.04$ (such a lower value compatible with \citep{2004MNRAS.351..169M}). Given these values, the relevance of the results here presented for the accretion efficiencies onto the compact DM cores becomes clear: depending on the compacity of the fermion core, the efficiency can go from  below the typical Schwarzschild case up to values larger than $0.1$, typical of Kerr BHs, as preferred by AGN population analysis. Further discussions about the challenges to observationally confirm the Kerr BH space-time at galaxy centers, mainly related to   
degeneracy of models regarding the mass of the compact object and, in particular, the spin parameter and possible deviations
from the Kerr metric, can be found in  \citet{bambi2017}.

%\noindent \textcolor{blue}{This results in the following values of $R_{\rm in}$ for the differnt models: agregar los valores aca o en otra tablita}

\begin{table*}%[ht]
    \caption[]{$r_{\rm in}$ considered for the different models.}
   	\label{table_2}
   	\centering
\begin{tabular}{lcccccccc}
\hline\hline %
Model & Particle mass  & $M_{\rm c}$ & $r_{\rm c}$ & $r_{\rm in}$ & $\varepsilon$\\ [0.01cm]
 & [keV]  & [$M_{\odot}$] & [$r_{\rm g}$]&  [$r_{\rm g}$] & [$\%$]\\ [0.01cm]
\hline
A1 & 48     &$1.5 \times 10^6$ &  $1.5\times 10^4$& $9.6\times 10^2$& 0.02\\ [0.01cm]
A2 & 48     &$1.0 \times 10^7$ & $1.0\times 10^3$ &$8.3\times 10^1$ & 0.2\\ [0.01cm]
A3 & 48     &$1.2 \times 10^8$ &  $3.4\times 10^1$& $2.6\times 10^0$ & 6.7\\ [0.01cm]

B1 & 200    &$3.5 \times 10^6$ & $8.1 \times 10^1$& $6.8 \times 10^0$ & 2.5\\ [0.01cm]
B2 & 200    &$1.0 \times 10^7$ &  $1.5 \times 10^1$& $1.3\times 10^0$ & 14\\ [0.01cm]

\hline\hline %
%\hline  \\[0.005cm]
\end{tabular}
\end{table*}

\subsection{The steady standard disc model embedded in DM} \label{3.2}

In this section we consider the model of steady thin discs developed by Shakura \& Sunyaev \citep{shakura1973}, and extend it to be applied in the context of the RAR-solutions. To this end, and following the treatment presented in \citet{frank2002}, we use cylindrical polar coordinates $(r, \phi, z)$, and assume that matter is very close to the plane $z=0$, and it is rotating with an angular velocity $\Omega(r)$  that remains very close to the Keplerian value:
\begin{equation}
    \Omega = \Omega_K(r) = \left(\frac{GM(r)}{r^3}\right)^{1/2}.
    \label{eq:Omega}
\end{equation}

\noindent Here, $M(r)$ is the mass distribution of the DM \textit{core}--\textit{halo} solution contained up to $r$. %The inner boundary condition taken into account is that the angular velocity in the disc around the compact object with mass  and radius $R_*$, until the accreting matter enters a boundary layer of radial extent b just outside the surface $r = R_*$ of the accreting star. Withinthis boundary layer $\Omega$ must decrease from a value $\Omega(R_* + b) \approx \Omega_K(R_* + b$) to the surface angular velocity $\Omega_* < \Omega_K(R_*)$. So that, 
The circular velocity is given by $v_\phi = r\Omega_K(r)$.

In addition, the gas is assumed to possess a small radial ‘drift’ velocity $v_R(r)$, which is negative near the central object, so that matter is being accreted. %It will be a function both of the radius and the time, as it would be desired to treat time-varying situations. 
The disc is characterised by its surface density $\Sigma(r)$, which is the mass per unit surface area of the disc, given by integrating the gas density $\rho$ in the $z$-direction. For a steady disc, the conservation of mass and angular momentum can be written as:% \citep{frank2002}:

\begin{equation}
     \Dot{M} = 2\pi r \Sigma (-v_R),
     \label{ratemass}
 \end{equation}
 
% \begin{equation}
%    R\Sigma v_R R^2 \Omega = \frac{G}{2\pi}+\frac{C}{2\pi},
%    \label{conser_ang_estacionario}
%\end{equation}

%\noindent respectively. Here, $\Dot{M}$ is the accretion rate (in units of g s$^{-1}$), $G(r)$ is the torque exerted by the inner ring on the exterior ring, and is given by:

% \begin{equation}
%     G(R) = 2\pi R\nu \Sigma R^2 \frac{d\Omega}{dr},
%     \label{torquesch}
% \end{equation}

%\noindent $\nu$ is the cinematic viscosity, and the constact $C$ is related to the rate of angular momentum to the compact object.

%\begin{equation*}
%    -\nu \Sigma \Omega^{'} = \Sigma (-v_R)\Omega + C/(2\pi R^3).
%\end{equation*}

%Combining Eqs. \ref{ratemass}-\ref{torquesch}, and taking into account that the mass of the central compact object is a function $M(r)$, we obtain:
%it is obtained:
\begin{equation}
    \eta\Sigma = \frac{\Dot{M}}{3\pi}\left[1-\left(\frac{M_{\rm in}r_{\rm in}}{M(r)r}\right)^{1/2}\right]
    \left[1-\frac{r}{3M(r)}\frac{dM(r)}{dr}\right]^{-1},
    \label{nusigma}
\end{equation}

\noindent respectively. Here, $\Dot{M}$ is the accretion rate (in units of g s$^{-1}$), $\eta$ is the cinematic viscosity,
%where $\nu$ is the viscosity, 
$r_{\rm in}$ is the internal radius of the disc (see section \ref{3.1}), and $M_{\rm in}=M(r_{\rm in})$. %We discuss the value of $R_{\rm in}$ in Sec. REF.

%[CHARLY: Aqui es importante decir de manera explicita que para llegar a esta solucion se tuvo que imponer la condicion de contorno (en analogia al caso del BH) que los torques viscosos se anulan en $R_{\rm in}$, ya que se asume que la estructura de disco Kepleriano como tal breaks down allí (ver footnote 2)]. 
Analogously as in the BH case, we adopt the viscous torques to vanish at the maximum of the binding energy, which in the case of the RAR model correspond to the inner radius $r_{\rm in}$. Thus it allows to explicitly obtain the viscous dissipation per unit disc face area as:
%Debido a que los torques viscosos generan \textit{disipación viscosa}, esta energía se irradia sobre las caras superior e inferior del disco, resultando en una disipación viscosa por unidad de área dada por:
\begin{equation}
\begin{aligned}
    D(r) &= %\frac{G}{4\pi R} \frac{d\Omega}{dr} =
    \frac{1}{2}\eta \Sigma \Big(r \frac{d\Omega}{dr} \Big)^2&\\
    &=\frac{3\Dot{M}}{8\pi}\frac{GM(r)}{r^3}
    \left[1-\left(\frac{M_{\rm in}r_{\rm in}}{M(r)r}\right)^{1/2}\right]%&\\
    %&& \times 
    \left[1-\frac{r}{3M(r)}\frac{dM(r)}{dr}\right].
    \label{Destandar}
    \end{aligned}
\end{equation}

%\noindent which results in:

%\begin{equation*}
 %   D(R) = \frac{3\Dot{M}}{8\pi}\frac{GM(R)}{R^3}\left[1-\left(\frac{M_{\rm in}R_{\rm in}}{M(R)R}\right)^{1/2}\right]
%\end{equation*}
%\begin{equation}
%     ~~~~~~~~~~~~~~~~~~~~~~~~~~~~~~~~~~~~~~~~~~~~~~~~~~~~  \left(1-\frac{R}{3M(R)}\frac{dM(R)}{dR}\right).
%\end{equation}

\noindent It can be seen that, as in the standard solution,  the viscous dissipation is independent of the physical nature of the viscosity $\eta$. %The other disc properties do depend on $\nu$.
Finally, the total disc luminosity is obtained by integrating $D(r)$ along the disc area.
%\begin{equation}
%    L_{\rm disc} = 2\int_{R_{\rm in}}^{\infty}D(R)2\pi R dR~~.
%\end{equation}

\begin{figure}
\centering
\includegraphics[width=80mm]{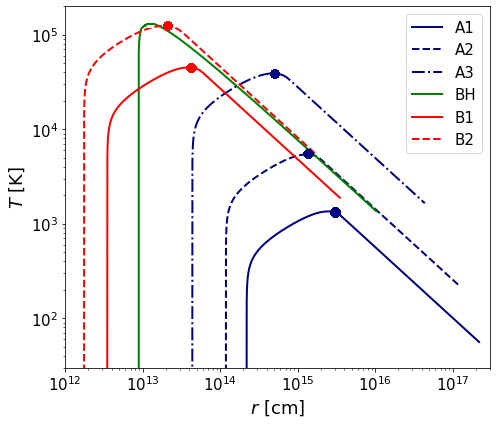}
\caption{Comparison of disc temperatures for the A and B 
RAR models with that of a BH of mass $M_{\rm BH} \sim 10^7 M_\odot$. The dots indicate the location of the $R_{\rm c}$ for each RAR model.}%same mass fermion \textcolor{red}{Incluir los Bs tmb. En la escala y se puede ir desde 1e30 hasta donde esta, y en la x desde 1e9 hasta donde esta, y asi se pueden agregar algunos tics mas sin sobrecargar todo} ($mc^2=48keV$) and different core masses.}
\label{fig:Temperature}
\end{figure}

We also consider an optically thick disc in the $z$-direction, hence each element of the disc radiates as a black-body with temperature $T(r)$, given by the equation of the viscous dissipation per unit disc face area  $D(r) = \sigma T^4(r)$, with $\sigma$ the Stefan-Boltzmann constant. In Fig. \ref{fig:Temperature} we show the comparison between the A and B RAR models, with a BH of $M \sim 10^7 M_\odot$.

For an observer at a distance $d$ from the centre of the disc whose line of sight makes an angle $i$ with respect to the normal of the disc plane, the flux at frequency $\nu$ is:
\begin{equation}
    F_\nu = \frac{4\pi h \cos i \nu^3}{c^2 d^2}\int_{r_{\rm in}}^{r_{\rm out}} \frac{rdr}{e^{h\nu/kT(r)}-1} .
    \label{flujo}
\end{equation}

%with the blackbody assumption. %At first, it can be discussed which is the appropiate $R_{\rm out}$ to consider.

%[CARLOS: agregar los 2 main statements que chequeamos respecto a la eleccion de $R_out$, es decir, el check de cuando se hace autogravitante, y el hecho de lo poco que varian los resultados al agrandarlo un orden de magnitud. Leer tambien el paper de Denimara Dias dos Santos et al, Apj 2023]\\

%In the following we set how the external edge condition of the disc is determined. According to \cite{BODGAN}, 
The outer radius $r_{\rm out}$ can be estimated using the condition that the disc becomes locally self-gravitating \cite{2015ApJ...800..124B}. This is determined by analysing the stability criterion for a differential rotation disc:

\begin{equation}
    Q_T = c_S\Omega/\pi G\Sigma >> 1,
\end{equation}

\noindent where $\Omega$ is the angular velocity given in Eq. \ref{eq:Omega}. The condition $Q_T = 1$ defines the self-gravitating disc:
\begin{equation}
    r_{\rm out} = (M/\pi \rho)^{1/3}.
    \label{rout}
\end{equation}
Knowing the density of the accretion disc, the outer radius can be determined. Further details are discussed in Appendix \ref{appen}, where we obtained $r_{\rm out} \approx 10^3~r_{\rm g}- 10^4~r_{\rm g}$, depending on the model studied. Nevertheless, the disc beyond $10^3~r_{\rm g}$ does not contribute significantly to the total luminosity, hence we adopt this value as the outer limit for all models.%As at this radius the disc becomes self-gravitating, varying $r_{\rm out}$ one order of magnitude do not impact on the flux profile obtained from Eq. \ref{flujo}. Moreover, it has been studied that the maximum at which the flux occurs do not impact on the spectrum analyzed when varying $r_{\rm out}$ from $\approx 10^3~r_{\rm in}$ to $r_{\rm out} \approx 10^4~r_{\rm in}$, .

\begin{figure}
\centering
\includegraphics[width=80mm]{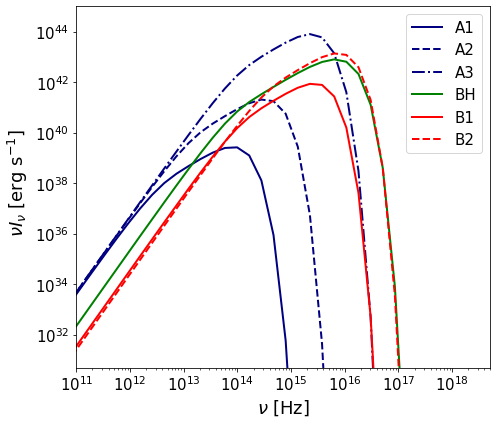}
\caption{Comparison of the luminosity for the A and B models with a BH of $M \sim 10^7 M_\odot$. The larger the mass of the DM-core, the more compact and dense is the degenerate core, implying the luminosity peak shifts towards higher $\nu$.}%same mass fermion \textcolor{red}{Incluir los Bs tmb. En la escala y se puede ir desde 1e30 hasta donde esta, y en la x desde 1e9 hasta donde esta, y asi se pueden agregar algunos tics mas sin sobrecargar todo} ($mc^2=48keV$) and different core masses.}
% Bottom: Luminosity for different DM-core masses of RAR models with same particle mass of $mc^2=48$ keV.
\label{fig:luminosity_ferrarese}
\end{figure}

In Fig. \ref{fig:luminosity_ferrarese} we show the luminosities obtained for models A and B. In all cases we have considered $\dot{M}=0.1\dot{M}_{\rm Edd}$, where $\dot{M}_{\rm Edd}$ is the Eddington accretion rate, defined as the accretion rate at which the compact source radiates at an efficiency of $\epsilon \sim 0.1$ of the Eddington luminosity, that is $\dot{M}_{\rm Edd} = L_{\rm Edd}/\epsilon c^2 \sim 1.4 \times 10^{16} \Big( M_{\rm c}/M_{\odot}\Big)$ erg s$^{-1}$.
% models, that is, different compact core masses for $48$ keV and $200$ keV fermions. 
Moreover, we show the comparison with a black hole of $M \sim 10^7 M_\odot$. See also points (ii) and (iii) in Section \ref{conclusions} for a relevant discussion regarding the RAR model predictions for disc luminosities.  % the comparison between the RAR model and Schwarzschild.

\begin{figure}
\centering
\includegraphics[width=80mm]{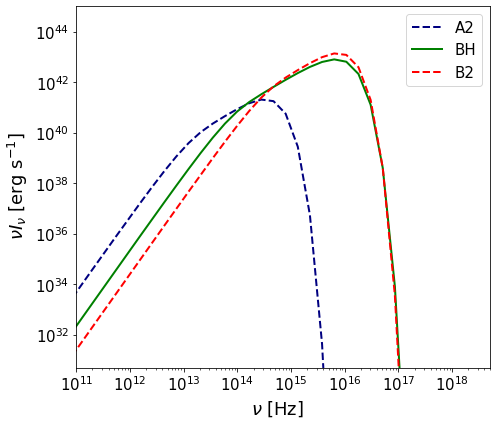}
\caption{Comparison of luminosity for the A2, B2 models and a BH of $\sim 10^7 \mathrm{M_\odot}$. Thus it exist a specific core compacity which produces a luminosity spectrum which is basically indistinguishable from that of a BH of the same mass as the DM core.}%same mass fermion ($mc^2=48keV$) and different core masses.}
\label{fig:luminosity_10e7Msun}
\end{figure}

Figure \ref{fig:luminosity_10e7Msun} only shows the luminosity for RAR models with $M_{\rm c} = 10^7 M_{\odot}$, using two values for the fermion mass: $mc^2=48$ keV (model A2) and $mc^2=200$ keV (model B2); we also compare this result with a disc around a Schwarzschild black hole of $M_{\rm BH} = M_{\rm c}$. This important result shows that it exist a core compacity (i.e. solution B2) for which the luminosity spectrum is basically indistinguishable from that of a Schwarzschild BH of the same mass as the DM core.

Different hypothesis are behind the standard thin disc approximation: the azimuthal velocity $v_\phi$ remains close to the Keplerian value, the disc remains thin at all radii, i.e. the height scale $H$ is much smaller than the extent of the disc $H<<r$, and the disc is optically thick in the $z$-direction. In order to corroborate if the solutions here obtained for the RAR model are consistent with the above thin disc ansatz, we consider a disc in hydrostatic equilibrium in the $z$-direction, meaning that there is no flow in the vertical direction. Then, for  $H<<r$ and $P\approx \rho c_s^2$ (with $c_s$ the sound speed), the solution satisfies:

%if the structure of the disc is considered to be in the z-direction, with no flow in that direction, the  can be written. Taking into account the 'thin' disc approximation implies $H<<r$ and $P\approx \rho c_s^2$. So, it is found for the RAR model:
\begin{equation}
    c_s <<  \left(\frac{GM(r)}{r}+G\frac{dM(r)}{dr}\right)^{1/2} \approx \left(\frac{GM(r)}{r}\right)^{1/2}~~.
    \label{c_s} 
\end{equation}
The local standard Kepler velocity should be highly supersonic.

\begin{figure}
    \centering
    \includegraphics[width=\linewidth]{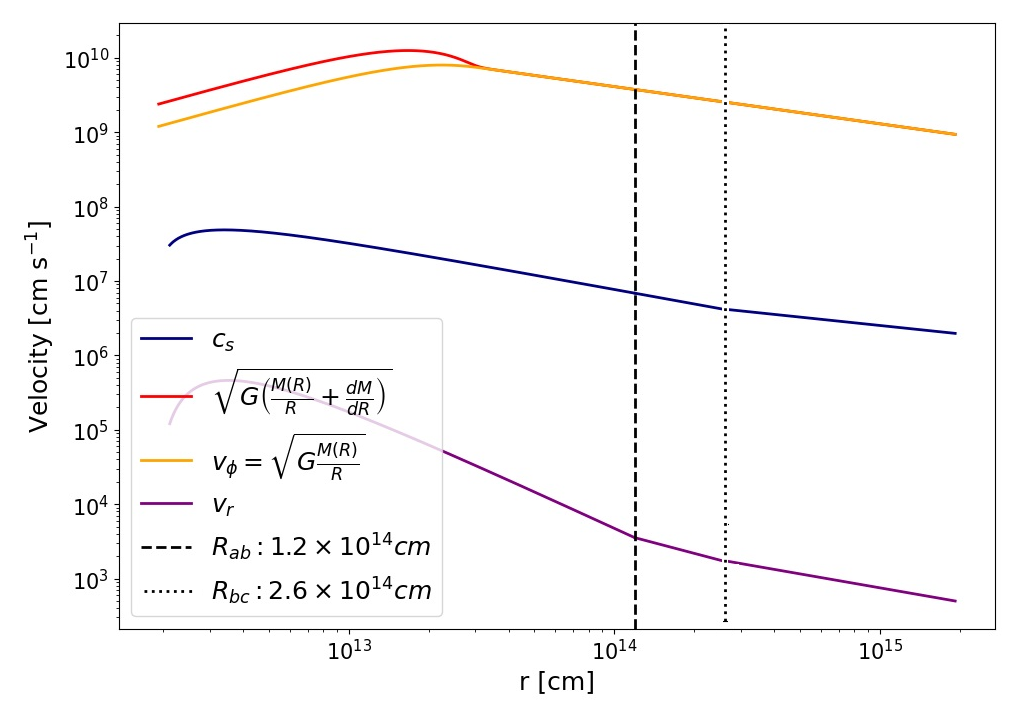}
    \caption{Hypothesis validation for model A2: Keplerian velocity must be highly supersonic, while radial velocity must be highly subsonic. Vertical dashed (and doted) lines specify $\rm R_{ab}$ (and $\rm R_{bc}$), i.e. the radii at which the behaviour of the disc changes between the inner and intermediate region (and between the intermediate and outer region) respectively (see Appendix \ref{appen}.) \\
    }
    \label{fig:cs}
\end{figure}

If the thin disc condition in Eq. \ref{c_s} holds, the circular matter velocity $v_\phi$ satisfies
\begin{equation}
    v_\phi = \sqrt{\frac{GM(r)}{r}+G\frac{dM(r)}{dr}}[1+O(M^{-2})].
\end{equation}
\noindent  
It can be seen that $v_\phi$ is very close to the Keplerian value, as we have assumed. Figure \ref{fig:cs} shows both hypothesis for model A2 as an example, though it is satisfied by all solutions as we have verified.  

%During the second part of the work, the discs solutions where analyzed. At first, in figure \ref{fig:vphi} it's shown the validity of the first hypotesis taken: that the circular veloicty must be close to the Keplerian value. In figure \ref{fig:cs}, the second assumption validity is shown: that the local Kepler velocity is highly supersonic. 

At this point it is worth to emphasise that the complete structure of the disc was not needed in order to compute the spectra for the RAR solutions. However, in order to be able to verify the hypothesis above mentioned, in Appendix A we write down the generalisation of the Shakura $\&$ Sunyaev set of equations within the fermionic DM model, and solve for the complete disc structure within the $\alpha$ prescription for the viscosity (see \cite{frank2002} for the analogous demonstration in the case of the standard $\alpha$ disc theory).

\section{Discussion and final remarks}\label{conclusions}

We have extended the standard steady thin disc model in order to study the accretion onto horizonless dark compact objects at galaxy centres. The BH alternative here investigated consists in a dense and highly degenerate core made of neutral fermions, surrounded by a more diluted mass distribution that is able to explain the DM halo in galaxies. Such a \textit{dense core}--\textit{diluted halo} DM configuration is known as the RAR model, and is a non-analytic solution of the Einstein equations of GR, which naturally arises once the quantum nature (i.e. Pauli principle) of the fermions is dully accounted for. The attention in this work was centred in active-like galaxies together with their central accretion processes, luminosity spectra and efficiencies. 

We have used core-halo RAR solutions for two different fermion masses, $48$ keV and $200$ keV, the later corresponding to a very dense DM-core of mass $10^7 M_\odot$, close to its critical value of gravitational collapse to a BH. The fact that the larger $m$ the more compact is the DM-core of given mass (see Fig. \ref{fig:m-rho}), makes the above choice particularly relevant to analyse how similar to a BH the luminosity spectra of $\alpha$-discs can be. Moreover, a particle mass falling within this range is totally compatible with both, linear structure formation in cosmology as well as non-linear structure formation, including galaxy rotation curves and scaling relations (see \cite{2023Univ....9..197A} for a review and references therein).  
%motivated by the successful application of the RAR model to: (i) the Milky Way, where the core-halo solutions can accurately reproduce the orbital motion of the S stars including its relativistic effects together with the outer halo \citep{bec2020,bec2021,arguelles2022sga}; (ii) different galaxy types from dwarf to elliptical, up to galaxy clusters; (iii) formation mechanism of supermassive BHs 

%In next we summarize t
The main results of our work can be summarised as follows: 

\begin{enumerate}
    \item The fact that the dense DM core is transparent implies the accretion disc can enter inside reaching down to event-horizon scales. As a consequence, it can achieve accretion efficiencies as large as $\epsilon=28.5\%$ and thus comparable to that of a highly rotating Kerr BH. In Fig. \ref{fig:binding} we have shown a relevant example comparing a $10^7 M_\odot$ Schwarzschild BH with two different RAR solutions of increasing particle mass ($A_2$ and $B_2$) having the same core mass. It is shown that for $m=48$ keV (solution $A_2$) the efficiency is below $1\%$ while for  $m=200$ keV (solution $B_2$) the stable (i.e. below critical) DM core is compact enough so to reach, at $r_{\rm in}<r_{\rm c}$, an accretion efficiency which is more than two times that of the BH. The maximum efficiency of $\epsilon=28.5\%$ is achieved for the critical core mass. The relevance of achieving Kerr-like efficiencies in active galaxies is supported by observational results \citep{2009MNRAS.396.1217R} based on the Soltan argument.
% Maximum efficiencies of the disc are reached at $r_{\rm in}<r_{\rm c}$, and they are found to be $ \sim 0.02 \% - 28.5 \%$, depending on the mass of the fermion. For greater masses (that is more compact solutions), the efficiency of the disc can be as high as the case of accretion onto black holes. In particular, for $mc^2 = 378$ keV, the efficiency of $28.5 \%$ is considerably higher to that of the accretion onto a Schwarzschild black hole of $5.7\%$.    
    \item At given DM core mass $M_c$ , the larger the particle mass the more compact is the degenerate core, and consequently the more luminous and energetic are the discs. This result, when considered in light of the efficiency trend explained above, implies it always exist a given core compacity producing a luminosity spectrum which is almost indistinguishable from that of a Schwarzschild BH of the same mass as the DM core. This important RAR model prediction is explicitly shown in Fig. \ref{fig:luminosity_10e7Msun} for a BH mass of $10^7 M_\odot$ typical of an active galaxy.
 
    \item At fixed DM particle mass it is possible to have different DM core masses $M_c$ (see Table \ref{table_1}) with surrounding DM halos fulfilling with the observationally inferred Ferrarese scaling relation \cite{ferrarese2002} (see Fig. \ref{fig:Ferrarese} and \cite{arguelles2019}). A novel RAR model prediction found in this work is that when $M_c$ increases, the peak frequency of the luminosity spectra also increases (see e.g. the trend in the red curves $B_1$ and $B_2$ of Fig. \ref{fig:luminosity_ferrarese}). This is due to the degeneracy of the core, which implies more compact solutions for more massive cores (at given $m$). This results is opposite to what is expected from accretion into a BH, and could be a fundamental tool to test the RAR model (see discussion below).   
    
\end{enumerate}

 Further detailed work is needed when trying to differentiate both paradigms of supermassive compact objects, as for example making use of real spectral energy distributions (SED) of AGNs, or to calculate the relativistic images produced by the emitted photons via ray tracing techniques. The later is an important project already started by our team, which will allow us to contrast with observations the shadow-like features predicted by the RAR model in contrast with the ones predicted by the BH.

Regarding the observational access to SEDs of AGNs relative to this work, the most important is the narrow window of low central object masses $\sim 10^6$-$10^7 M_\odot$ (for the $m=200$ keV case),  or $\sim 10^7$-$10^8 M_\odot$ (for $m=48$ keV),  before the corresponding DM core becomes critical and collapses to a BH \footnote{Once the SMBH is formed from DM core collapse it will growth further via baryonic accretion \citep{2023MNRAS.523.2209A} having the standard spectral energy features.}. An eventual observational detection of luminosity peaks which shifts towards higher $\nu$ in such a small DM core-mass window is challenging. This is mainly because of the lack of data in the UV band (at about the UV bump) due to the absorption of the IGM; the large error bars at about the blue bump;  or that most of observed SEDs are obtained for relatively large central object masses $> 10^8 M_\odot$ \citep{2017MNRAS.465..358C}.  

All in all, the original results presented in this work and summarised above may imply an important landmark -and may open new avenues of research- in the field of AGN theory and phenomenology in connection to DM physics and SMBHs.

\begin{acknowledgements}
    C.R.A. acknowledges supported from CONICET of Argentina, the ANPCyT (grant PICT-2018-03743), and ICRANet. F.L.V. acknowledges support from the Argentine agency CONICET (PIP 2021-0554). V.C. thanks financial support from CONICET, Argentina. MFM acknowledges support from CONICET (PIP2169) and from the Universidad Nacional de La Plata (G168).
\end{acknowledgements}

   \bibliographystyle{aa} % style aa.bst
   \bibliography{ref} % your references Yourfile.bib
%%%%%%%%%%%%%%%%% APPENDICES %%%%%%%%%%%%%%%%%%%%%
\begin{appendix}
\section{Local structure of the disc}\label{appen}

%\subsection{Local structure of the discs}%{$\alpha$-discs}

We present here the solution of the complete disc structure. To this end, we consider, as in the standard disc solution, the $\alpha-$ prescription for the viscosity, given by:
\begin{equation}
\eta = \alpha c_s H.
\end{equation}

The main objective of the following theoretical content of this Appendix is done both, for completeness, and in order to be able to verify the three main hypothesis of the (extended) Shakura $\&$ Sunyaev theory: the geometrically thin disc ($H(r)<< r$)  and the  Keplerian velocity approximation, together with the optically thick ($\tau > 1$) assumption, for which the solution of the full disc structure is needed. An example of such a verification was shown in Fig. \ref{fig:cs} of the main text.

In the thin disc approximation, the determination of the disc structure is simplified. Following \citet{shakura1973,frank2002}, and using the results obtained in Section \ref{3.2} for the generalised thin disc embedded within the RAR DM distribution, the set of disc equations results in:
  
\begin{enumerate}
     \item $\rho = \Sigma/H$, \label{1r}
    \item H = $c_s\left(\frac{GM(r)}{r}+G\frac{dM(r)}{dr}\right)^{-1/2}r$,\label{2r}
    \item $c_s^2=P/\rho$, \label{3r}
    \item $P=\frac{\rho(r)kT_c}{\mu m_p} + \frac{4\sigma}{3c}T_c^4$,\label{4r}
    \item $\frac{4\sigma T_c^4}{3\tau} = \frac{3\Dot{M}}{8\pi}\frac{GM(r)}{r^3}\left[1-\left(\frac{M_{\rm in}r_{\rm in}}{M(r)r}\right)^{1/2}\right]\left(1-\frac{r}{3M(r)}\frac{dM(r)}{dr}\right)$,\label{5r}
    \item $\tau = \Sigma \kappa_R(\rho,T_c)=\tau(\Sigma,\rho,T_c) $,\label{6r}
    \item  $\eta\Sigma = \frac{\Dot{M}}{3\pi}\left[1-\left(\frac{M_{\rm in}r_{\rm in}}{M(r)r}\right)^{1/2}\right]\frac{1}{1-\frac{r}{3M(r)}\frac{dM(r)}{dr}}$,\label{7r}
    \item $\eta = \eta(\rho,T_c,\Sigma,\alpha,\dots)$.\label{8r}
    \label{ecs}
\end{enumerate}

\noindent It is worth mentioning that in the limits $M(r)\rightarrow M$ and \mbox{$\frac{dM(r)}{dr}\rightarrow 0$}, the standard disc solutions are recovered.

%To compute the local structure of the disc, it is necessary to adopt a viscosity function. 

%\begin{equation*}
%    \eta = \alpha c_s H.
%\end{equation*}

There are three distinct regions in the disc, determined by the relevant absorption mechanism, and the importance of $P_{\rm rad}$ versus $P_{\rm gas}$. Regarding the absorption mechanisms, we consider two main processes: free-free absorption, where the opacity is given by (see e.g. \cite{shakura1973})
\begin{equation}
    \kappa_{\rm ff} = 5.0 \times 10^{24} \rho \textrm{T}_c^{-7/2}\textrm{~cm}^2\textrm{g}^{-1},
\end{equation}

\noindent and scattering, where we adopt:

\begin{equation}
   \kappa_{\rm sc} = \frac{\sigma_{\rm T}}{m_p} \sim 0.4 \textrm{~cm}^2\textrm{g}^{-1}.
\end{equation}

The parameterization of the complete solutions, together with the regions where $P_{\rm gas}$ dominates over $P_{\rm rad}$ (and vice versa) can be found in Appendix \ref{appen}.
 
%In the outer region, the gas pressure dominates over the radiation pressure, $P_{\rm gas} >> P_{\rm rad}$, and the opacity is mainly due to free-free absorption. We use the opacity given by (REF) xx 

%and assumed that $\rho$ and $T_c$ are such that the Rosseland mean opacity is approximated by Kramers’ law:
%\begin{equation*}
%    \kappa_R = 5 \times 10^{24} \rho \textrm{T}_c^{-7/2}\textrm{cm}^2\textrm{g}^{-1}.
%\end{equation*}

%The radiation pressure term from the equation of state can be dropped. This assumption will be checked with the disc solution, in order to check if it is valid. 
 %Finally, this implies that the $\alpha$ value can not be measured by observable measurements of the steady-disc.

%\item Intermediate region: $P_{\rm gas}>>P_{\rm rad}$ y $\kappa_{es}>>\kappa_{ff}$:

%\item Inner region: Para la región interna (a), donde $P_{\rm rad}>>P_{\rm gas}$ y $\kappa_{es}>>\kappa_{ff}$:

As in the standard solution, there are three distinct regions in the disc: (a) an inner region, where $P_{\rm rad} \gg P_{\rm gas}$ and $\kappa_{es} \gg \kappa_{ff}$, (b) an intermediate region, where $P_{\rm gas} \gg P_{\rm rad}$ and $\kappa_{es} \gg \kappa_{ff}$, and (c) an outer region, where $P_{\rm gas} \gg P_{\rm rad}$ and $\kappa_{es} \ll \kappa_{ff}$. The complete solution of the set of Eqs. \ref{1r}-\ref{8r} for the inner, intermediate and outer region are presented in Eqs. \ref{in_i}-\ref{in_u}, \ref{int_i}-\ref{int_u} and \ref{eq1}-\ref{eq7}, respectively. We have defined $R_{10}= r/(10^{10}\textrm{ cm})$, $m_1 =M(r)/M_{\odot}$, $\Dot{M}_{16} = \Dot{M}/(10^{16} \textrm{g s}^{-1})$, $f^4 = 1-\left(\frac{M_{\rm in}r_{\rm in}}{M(r)r}\right)^{1/2}$, $A=\left(\frac{M(r)}{r}+\frac{dM}{dr}\right)$, $B=1-\frac{r}{3M(r)}\frac{dM(r)}{dr}$ and we have considered $\mu = 0.615$ for a fully ionised gas.%, the disc structure results in Eqs. \ref{eq1}-\ref{eq7}.

It is worth noticing that the limits $\frac{dm_1}{dR_{10}} \rightarrow 0$ and $M(r)\rightarrow M_{\rm in}$ recover the standard disc solution around a compact object of given mass $M$. Secondly, the $\alpha$ power is of the same order of magnitude than that of the standard solution. Hence, as $\alpha$ powers are low, the magnitudes calculated for the disc are not sensitive to the value of $\alpha$.

The transition radii between the regions result in:

\begin{equation}
    %R_{ab} \approx (2.3\times 10^{-10})^{1/4}\alpha^{1/16}m_1^{1/2}f^{2}\Dot{M}_{16}^{1/2}B^{1/2}A^{-11/32}(10^{10}\mathrm{cm}), 
    R_{ab} \approx 3.9 \times 10^{7} \alpha^{1/16}m_1^{1/2}f^{2}\Dot{M}_{16}^{1/2}B^{1/2}A^{-11/32} \mathrm{~cm}, 
    \label{A1}
\end{equation}

%El radio para el cual cambia la región del disco:
\begin{equation}
    %R_{bc} \approx (4.87\times 10^{-10})^{1/2}m_1^{1/2}f^{2}\Dot{M}_{16}^{1/2}B^{1/2}A^{-1/4}(10^{10}\mathrm{cm}) 
    R_{bc} \approx 2.2\times 10^{5}m_1^{1/2}f^{2}\Dot{M}_{16}^{1/2}B^{1/2}A^{-1/4} \mathrm{~cm} 
    \label{A2}
\end{equation}

\begin{figure*}
\begin{flushleft}
Inner region of the disc: $P_{\rm rad} \gg P_{\rm gas}$ and $\kappa_{es} \gg \kappa_{ff}$:
\end{flushleft}
\vspace{-0.3cm}
\noindent\makebox[\linewidth]{\rule{\textwidth}{0.4pt}}
\begin{align}
      \Sigma &= 5.0\times 10^{8}\alpha^{-4/5}f^{12/5}B^{-2}Am_1^{1/5}\Dot{M}_{16}^{3/5}R_{10}^{-3/5}~\mathrm{g~cm^{-2}};\\
      H &= 1.4\times 10^{4} m_1^{-7/20}R_{10}^{21/20}\Dot{M}_{16}^{1/5}f^{4/5} \textrm{cm}; \label{in_i}\\
      \rho &= 3.7\times 10^{4}\alpha^{-4/5}f^{8/5}B^{-2}Am_1^{11/20}\Dot{M}_{16}^{2/5}R_{10}^{-33/20}~\mathrm{g~cm^{-3}};\\
     T_c &=7.7\times 10^{5}\alpha^{-1/5}m_1^{3/10}B^{-1/2}A^{1/4}R_{10}^{-9/10}\Dot{M}_{16}^{2/5}f^{8/5} K; \label{in_u}\\
    \tau &= 2.3\times 10^{17}\alpha^{-4/5}f^{12/5}B^{-2}Am_1^{1/5}\Dot{M}_{16}^{3/5}R_{10}^{-3/5};\\
    \eta &= 9.4\times 10^7 \alpha m_1^{13/20}A^{-1/2}R_{10}^{-19/20}\Dot{M}_{16}^{2/5}f^{24/5} \mathrm{cm^2~s^{-1}};\\
    v_R &= 9.4\times 10^{-3} m_1^{13/20}A^{-1/2}R_{10}^{-39/20}\Dot{M}_{16}^{2/5}f^{24/5} \mathrm{cm~s^{-1}}.
\end{align}
\vspace{-0.3cm}
%\noindent\makebox[\linewidth]{\rule{\textwidth}{0.4pt}}
%\end{figure*}
%\begin{figure*}
\begin{flushleft}
Intermediate region: $P_{\rm gas} \gg P_{\rm rad}$ and $\kappa_{es} \gg \kappa_{ff}$:
\end{flushleft}
\vspace{-0.3cm}
\noindent\makebox[\linewidth]{\rule{\textwidth}{0.4pt}}
\begin{align}
    \Sigma &=9.9\alpha^{-4/5}f^{12/5}m^{-1/5}\Dot{M}_{16}^{3/5}R_{10}^{-1/5}A^{2/5}B^{-1}~ \mathrm{g~cm}^{-2};\\     
     H &= 9.6\times 10^7\alpha^{-1/10}m_1^{1/10}\Dot{M}_{16}^{1/5}R_{10}^{3/5}f^{4/5}A^{-9/20} \mathrm{cm};\label{int_i}\\
     \rho &= 1.0\times10^{-7}\alpha^{-7/10}m_1^{-3/10}\Dot{M}_{16}^{2/5}R_{10}^{-4/5}f^{8/5}A^{13/20}B^{-1}~ \mathrm{g~cm}^{-1};\\
     T_c &= 9.2\times 10^3\alpha^{-1/5}M^{1/5}\Dot{M}_{16}^{2/5}R_{10}^{-4/5}f^{8/5}A^{1/10} \mathrm{K};\\
     \tau &= 6.9\times 10^4 \alpha^{-4/5}f^{-8/5}m_1^{-6/5}\Dot{M}_{16}^{-2/5}R_{10}^{9/5}A^{7/10}B^{-2}; \\
     \eta &= 4.3\times 10^{9}\alpha^{9/10}A^{-9/20}R_{10}^{3/5}m_1^{1/10}\Dot{M}_{16}^{1/5}f^{4/5}\mathrm{cm^2~s^{-1}};\\
     v_R &= 4.3\times 10^{-1}\alpha^{9/10}A^{-9/20}R_{10}^{-2/5}m_1^{1/10}\Dot{M}_{16}^{1/5}f^{4/5}\mathrm{cm~s^{-1}}.
     \label{int_u}
\end{align}
\label{inter}
%\noindent\makebox[\linewidth]{\rule{\textwidth}{0.4pt}}
%\end{figure*}
%\begin{figure*}
\begin{flushleft}
Outer region: $P_{\rm gas} \gg P_{\rm rad}$ and $\kappa_{es} \ll \kappa_{ff}$:
\end{flushleft}
\vspace{-0.3cm}
\noindent\makebox[\linewidth]{\rule{\textwidth}{0.4pt}}
\begin{align}
     \Sigma &= 3.65\alpha^{-4/5}
         A^{7/20
         }R_{10}^{-2/5}f^{14/5}
         B^{-9/10}
     \Dot{M}_{16}^{7/10}m_1^{-1/10} \mathrm{g~cm}^{-2}; \label{eq1}\\
     H &= 1.5\times10^{8}\alpha^{-1/10}f^{3/5}\Dot{M}_{16}^{3/20}R_{10}^{7/10}m_1^{1/20}
         A^{-17/40}
         B^{-1/20}\mathrm{cm};\label{eq2}\\
     \rho &= 2.4\times10^{-8}\alpha^{-7/10}R_{10}^{-11/10}f^{11/5}\Dot{M}_{16}^{11/20}m_1^{-3/20}
         A^{31/40}
        B^{-17/20}\mathrm{g~cm}^{-3};\label{eq3}\\
     T_c &= 2.5\times 10^{4} \alpha^{-1/5}R_{10}^{-3/5}\Dot{M}_{16}^{3/10}f^{6/5}m_1^{1/10}
        A^{3/20}
        B^{-1/10} \mathrm{K};\label{eq4}\\
     \tau &= 177\alpha^{-4/5}R_{10}^{3/5}f^{4/5}\Dot{M}_{16}^{1/5}m_1^{-3/5}
        A^{3/5}B^{-7/5};\label{eq5}\\
     \eta &= 2.7\times 10^{14}\alpha^{4/5}f^{6/5}\Dot{M}_{16}^{3/10}R_{10}^{2/5}m_1^{1/10}
        A^{-7/20}
       B^{-1/10}\mathrm{~cm}^2 \mathrm{~s}^{-1};\label{eq6}\\
     v_R &= 2.7\times 10^{4}\alpha^{4/5}f^{-14/5}\Dot{M}_{16}^{3/10}R_{10}^{-3/5}m_1^{1/10}
        A^{-7/20}
       B^{9/10}\mathrm{cm~s}^{-1}.
       \label{eq7}
\end{align}
\label{outer}
%\noindent\makebox[\linewidth]{\rule{\textwidth}{0.4pt}}
\end{figure*}

We now consider the case of models A2 and B2 described in Table \ref{tab: 1}. We assume $\Dot{M} = 0.1 M_{\rm Edd} \approx 1.4\times 10^{23}$ g~s$^{-1}$, and $\alpha =0.1$. We also compare to the standard disc around a Schwarzschild black hole of mass $M_{\rm BH}= 10^7 M_{\odot}$.

\begin{figure*}
\centering
\includegraphics[width=0.31\textwidth]{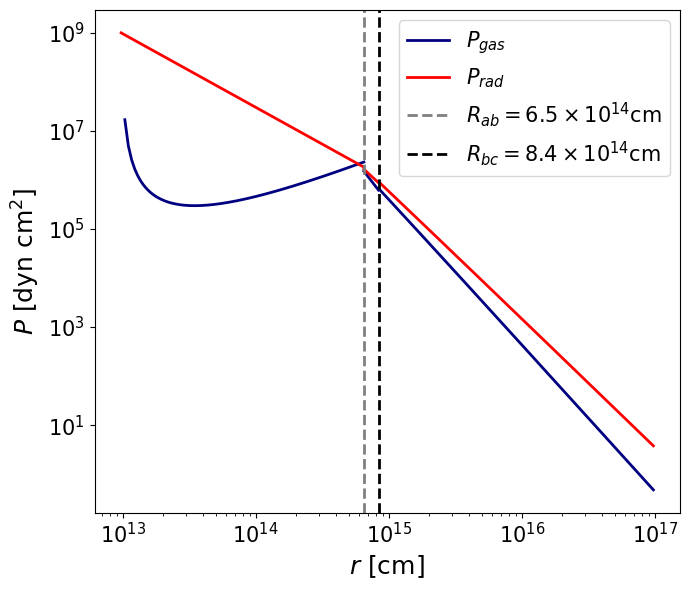}
\includegraphics[width=0.31\textwidth]{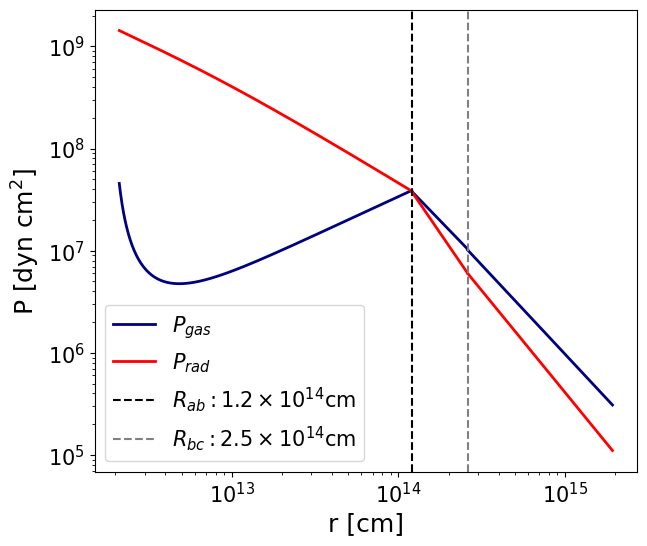}
\includegraphics[width=0.3\textwidth]{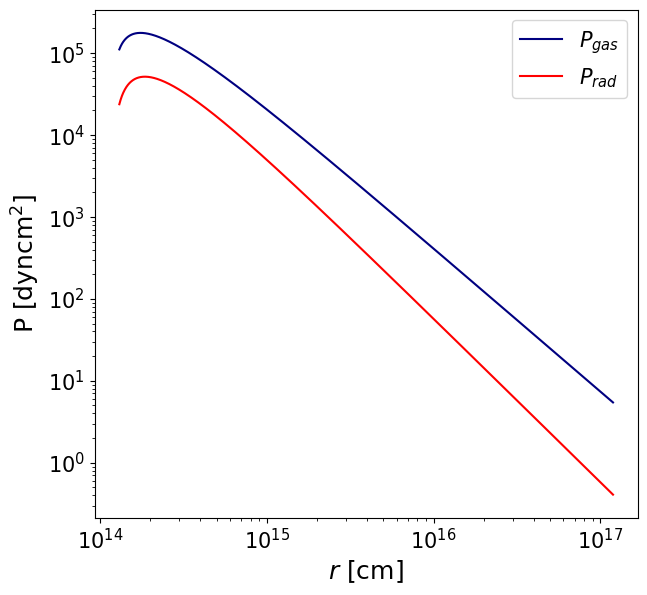}
\caption{Variation of the radiation and gas pressures for a disc around a black hole (left), the RAR solution B2 (centre) and the RAR solution A2 (right). Dashed black and grey lines represent the limits of the different regions.} \label{fig:presiones}
\end{figure*}

\begin{figure*}
\centering
\includegraphics[width=0.3\textwidth]{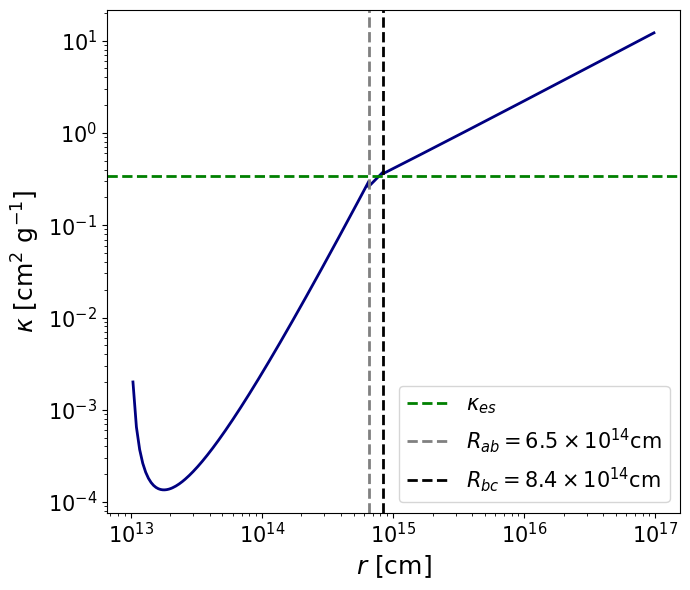}
\includegraphics[width=0.31\textwidth]{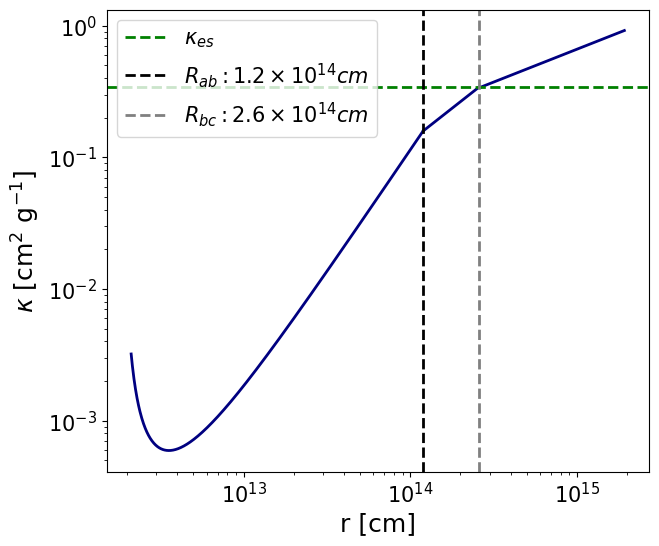}
\includegraphics[width=0.3\textwidth]{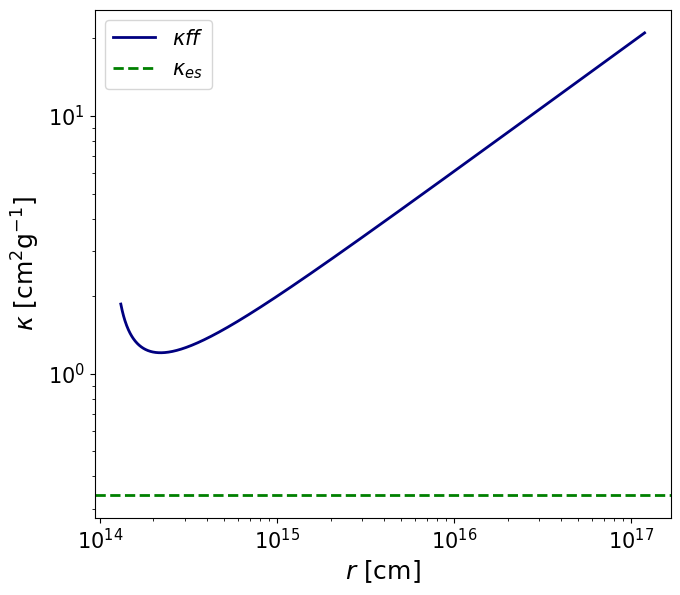}
\caption{Variation of the absorption coefficients for a disc around a black hole (left), the RAR solution B2 (centre) and the RAR solution A2 (right). Dashed black and grey lines represent the limits of the different regions.} \label{fig:opacidades}
\end{figure*}

In Fig. \ref{fig:presiones} we show the variation of both contributions to the total pressure in the different regions; in an analogous way, in  Fig. \ref{fig:opacidades} we show the absorption coefficients due to scattering and free-free absorption. In both cases, we have compared our models (centre and right) to the standard disc around a black hole (left).

% \textcolor{purple}{Valen: No pasa nada con las discontinuidades cuando pasa de region? Habria que hacer algun comentario en el texto sobre esto? Ya que los ejes $Y$ mas o menos van en el mismo rango por ahi estaria bueno que tengan este eje compartido. Habría que agrandar los labels.}

\begin{figure*}
\centering
\includegraphics[width=0.31\textwidth]{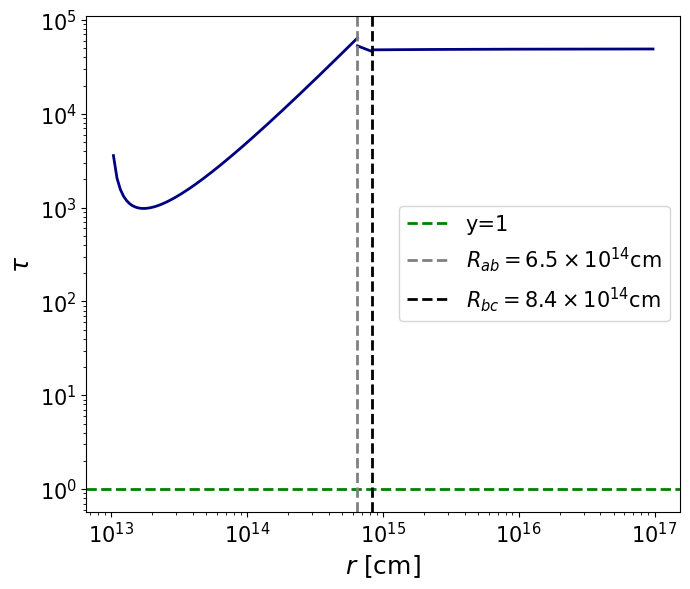}
\includegraphics[width=0.31\textwidth]{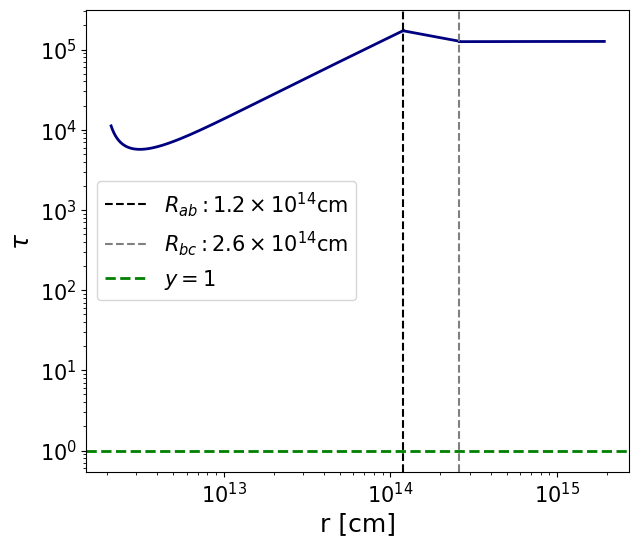}
\includegraphics[width=0.3\textwidth]{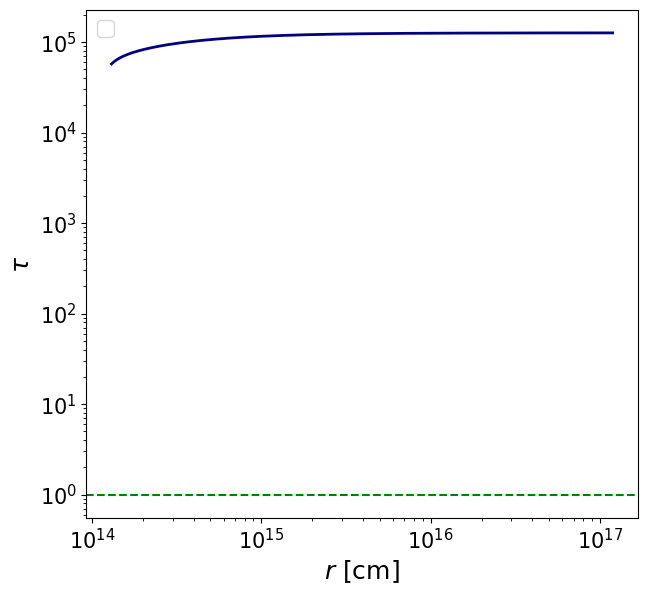}
\caption{Variation of the opacity for a disc around a black hole (left), the RAR solution B2 (centre) and the RAR solution A2 (right). Dashed black and grey lines represent the limits of the different regions.} \label{fig:tau}
\end{figure*}

\begin{figure*}
\centering
\includegraphics[width=0.3\textwidth]{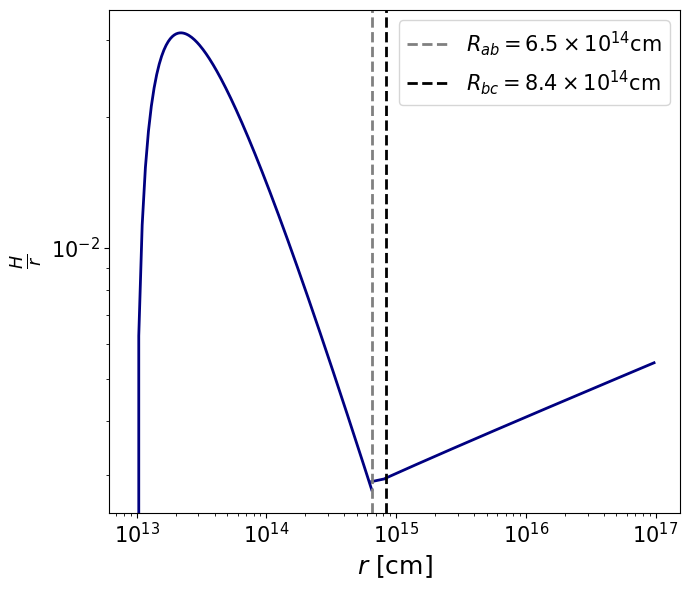}
\includegraphics[width=0.3\textwidth]{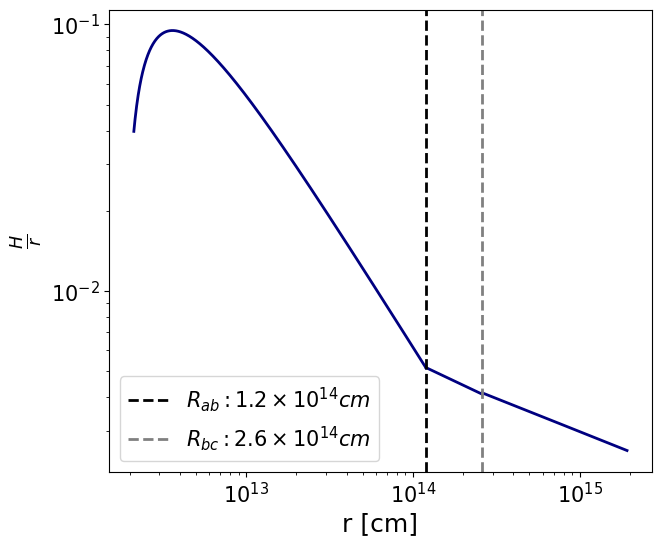}
\includegraphics[width=0.3\textwidth]{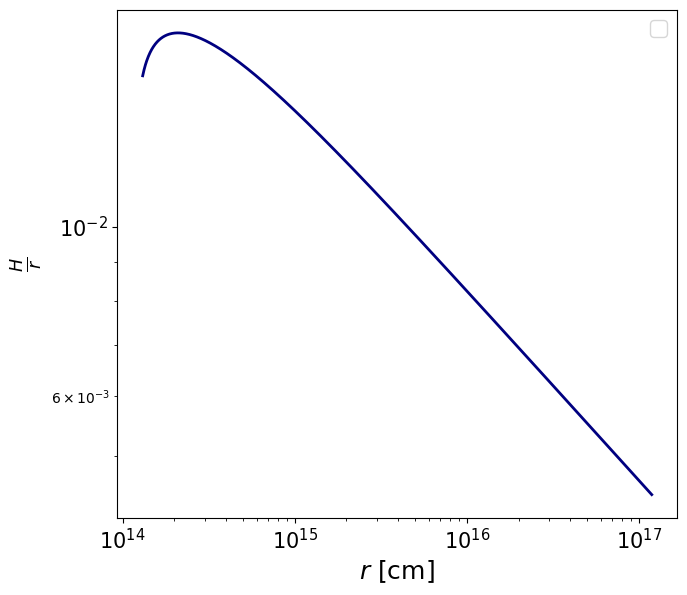}
\caption{Thin disc approximation for a disc around a black hole (left), the RAR solution B2 (centre) and the RAR solution A2 (right). Dashed black and grey lines represent the limits of the different regions.} \label{fig:geometrythindisk}
\end{figure*}

An interesting result is obtained varying the fermion mass from $200$ keV to $48$ keV (for the same mass of the core, model A2); in this case, only  the outer region exists. This is due to the fact that the compacity of the core diminish for lower masses of the fermion, which leads the disc to reach lower temperatures, as it can be seen in Fig. \ref{fig:Temperature}. 

Using the criterion given by Eq. \ref{rout}, we can determined the outer radius of the disc. For both, model B2 and the black hole, $r_{\rm out} = 1.9\times 10^{15}\mathrm{cm}$ and $r_{\rm out} = 1.0 \times 10^{14}\mathrm{cm}$, respectively, which implies for both cases $r_{\rm out} \approx 10^{3}~r_{\rm g}$. It is interesting to analyse model A2, where only the outer region of the disc is present. In this case, we found that the disc becomes self-gravitating at $r_{\rm out} = 1.3\times 10^{16}\mathrm{cm} \approx 10^4 r_{\rm g}$.
%the radius at which the inner and intermediate regions should change is $R_{ab} = 8.73\times 10^{14}\mathrm{cm}$. On the other hand, 
% for the inner region at $r_{\rm out} =  10^{14}\mathrm{cm}$. As $r_{\rm out} < R_{ab}$, the inner region do not exists. Same behaviour occurs with the intermediate region, happening to be $R_{bc} = 2.1\times 10^{15}\mathrm{cm}$ and $r_{\rm out} =  10^{15}\mathrm{cm}$. Finally, for the outer region is found that $r_{\rm out} = 1.3\times 10^{16}\mathrm{cm}$, bigger than $R_{bc}$, and so the outer region do exist. In this case 

%$r_{\rm out} \approx 10^2~r_{\rm in}$ {\bf idem comment anterior}.
%

%An interesting result is that, for the same mass of the core, we vary the fermion mass to  $56$ keV and $100$ keV we only obtain an outer region. This is due to the fact that the compacity of the core diminiss for lower masses of the fermion. (??).
%Es interesante comparar con una misma masa del núcleo co ese6} se muestra la opacidad en función del radio. A diferencia de lo obtenido para $mc^2=378\{keV}$, no se reproducen las tres regiones del disco, existiendo únicamente la región exterior, donde domina la opacidad free-free y la presión del gas. Dicho resultado se condice con el hecho de que la compacidad de los objetos para $mc^2=56\{keV}$ y $mc^2 = 100\{keV}$ es menor en comparación con $mc^2 = 378\{keV}$ 

\end{appendix}
\end{document}